\documentclass[aps,prl,twoside,twocolumn,floatfix,superscriptaddress,amsmath,showpacs,amssymb]{revtex4}
\usepackage{amssymb}
\usepackage{bm}  
\usepackage{graphicx}
\usepackage{psfrag}

\def\buch{Institute for Nuclear Physics and Engineering, Bucharest, Romania}
\def\buda{KFKI Research Institute for Particle and Nuclear Physics, Budapest, Hungary}
\def\cler{Clermont Universit\'{e}, Universit\'{e} Blaise Pascal, CNRS/IN2P3, Laboratoire de Physique Corpusculaire, BP 10448, F-63000 Clermont-Ferrand, France}
\def\sp{University of Split, Split, Croatia}
\def\darm{Gesellschaft f\"{u}r Schwerionenforschung, Darmstadt, Germany}
\def\dres{Institut f\"{u}r Strahlenphysik, Forschungszentrum Dresden-Rossendorf, Dresden, Germany} 
\def\heid{Physikalisches Institut der Universit\"{a}t Heidelberg, Heidelberg, Germany}
\def\mosc{Institute for Theoretical and Experimental Physics, Moscow, Russia}
\def\kurc{Kurchatov Institute, Moscow, Russia}
\def\seou{Korea University, Seoul, Korea}
\def\stra{Institut Pluridisciplinaire Hubert Curien, In2P3/CNRS, and Universit\'{e} de Strasbourg, Strasbourg, France}
\def\wars{Institute of Experimental Physics, University of Warsaw, Warsaw, Poland}
\def\zagr{Ru{d\llap{\raise 1.22ex\hbox{\vrule height 0.09ex width 0.2em}}\rlap{\raise 1.22ex\hbox{\vrule height 0.09ex width 0.06em}}}er Bo\v{s}kovi\'{c} Institute, Zagreb, Croatia}
\def\lan{Institute of Modern Physics, Chinese Academy of Sciences, Lanzhou, China}
\def\mun{Physik Department, Technische Universit\"{a}t M\"{u}nchen, D-85748 Garching, Germany}
\def\vien{Stefan-Meyer-Institut f\"{u}r Subatomare Physik, \"{O}sterreichische Akademie der Wissenschaften, Boltzmanngasse 3, A-1090, Wien, Austria}
\def\jap{Heavy-Ion Nuclear Physics Laboratory, RIKEN, Wako, Saitama 351-0198, Japan}

\begin{document}

\title[]{ Measurement of $K^*(892)^0$ and $K^0$ mesons in Al+Al collisions at 1.9$A$ GeV}

\author{X.~Lopez} \email{lopez@clermont.in2p3.fr} \affiliation{\cler}
\author{N.~Herrmann} \affiliation{\heid}
\author{K.~Piasecki} \affiliation{\heid} \affiliation{\wars}
\author{A.~Andronic} \affiliation{\darm}
\author{V.~Barret} \affiliation{\cler}
\author{Z.~Basrak} \affiliation{\zagr}
\author{N.~Bastid} \affiliation{\cler}
\author{M.L.~Benabderrahmane} \affiliation{\heid}
\author{P.~Buehler} \affiliation{\vien}
\author{M.~Cargnelli} \affiliation{\vien}
\author{R.~\v{C}aplar} \affiliation{\zagr}
\author{P.~Crochet} \affiliation{\cler}
\author{P.~Dupieux} \affiliation{\cler}
\author{M.~D\v{z}elalija} \affiliation{\sp}
\author{L.~Fabbietti} \affiliation{\mun}
\author{I.~Fija{\l}-Kirejczyk} \affiliation{\wars}
\author{Z.~Fodor} \affiliation{\buda}
\author{P.~Gasik} \affiliation{\wars}
\author{I.~Ga\v{s}pari\'c} \affiliation{\zagr}
\author{Y.~Grishkin} \affiliation{\mosc}
\author{O.N.~Hartmann} \affiliation{\vien}
\author{K.D.~Hildenbrand} \affiliation{\darm}
\author{B.~Hong} \affiliation{\seou}
\author{T.I.~Kang} \affiliation{\seou}
\author{J.~Kecskemeti} \affiliation{\buda}
\author{M.~Kirejczyk} \affiliation{\wars}
\author{Y.J.~Kim} \affiliation{\darm}
\author{M.~Ki\v{s}} \affiliation{\darm} \affiliation{\zagr}
\author{P.~Koczon} \affiliation{\darm}
\author{M.~Korolija} \affiliation{\zagr}
\author{R.~Kotte} \affiliation{\dres}
\author{A.~Lebedev} \affiliation{\mosc}
\author{Y.~Leifels} \affiliation{\darm}
\author{V.~Manko} \affiliation{\kurc}
\author{J.~Marton} \affiliation{\vien}
\author{T.~Matulewicz} \affiliation{\wars}
\author{M.~Merschmeyer} \affiliation{\heid}
\author{W.~Neubert} \affiliation{\dres}
\author{D.~Pelte} \affiliation{\heid}
\author{M.~Petrovici} \affiliation{\buch}
\author{F.~Rami} \affiliation{\stra}
\author{W.~Reisdorf} \affiliation{\darm}
\author{M.S.~Ryu} \affiliation{\seou}
\author{P.~Schmidt} \affiliation{\vien}
\author{A.~Sch\"{u}ttauf} \affiliation{\darm}
\author{Z.~Seres} \affiliation{\buda}
\author{B.~Sikora} \affiliation{\wars}
\author{K.S.~Sim} \affiliation{\seou}
\author{V.~Simion} \affiliation{\buch}
\author{K.~Siwek-Wilczy\'{n}ska} \affiliation{\wars}
\author{V.~Smolyankin} \affiliation{\mosc}
\author{K.~Suzuki} \affiliation{\vien}
\author{Z.~Tyminski} \affiliation{\wars}
\author{P.~Wagner} \affiliation{\stra}
\author{E.~Widmann} \affiliation{\vien}
\author{K.~Wi\'{s}niewski} \affiliation{\wars}
\author{Z.G.~Xiao} \affiliation{\lan}
\author{I.~Yushmanov} \affiliation{\kurc}
\author{X.Y.~Zhang} \affiliation{\lan}
\author{A.~Zhilin} \affiliation{\mosc}
\author{J.~Zmeskal} \affiliation{\vien}

\collaboration{FOPI Collaboration}
\noaffiliation

\date{\today}

\author{P.~Kienle }  \affiliation{\vien} \affiliation{\mun}
\author{T.~Yamazaki} \affiliation{\jap}

\begin{abstract}
New measurement of sub-threshold $K^*(892)^0$ and $K^0$ production is presented. 
The experimental data complete the measurement of strange particles produced in
Al+Al collisions at 1.9$A$ GeV measured with the FOPI detector at SIS/GSI.
The $K^*(892)^0$/$K^0$ yield ratio is found to be $0.0315\pm 0.006 (\mathrm{stat.})\pm 0.012 (\mathrm{syst.})$
and is in good agreement with the UrQMD model prediction. These measurements
provide information on in-medium cross section of $K^+$-$\pi^-$ fusion which is the
dominant process on sub-threshold $K^*(892)^0$ production.
\end{abstract}

\maketitle


The FOPI Collaboration has performed a high statistics experiment to study 
strangeness production in Al+Al collisions at a beam kinetic
energy of 1.9$A$ GeV. The collected data sample is large enough to enable
for the first time the reconstruction of deep sub-threshold resonance production
of $K^*(892)^0$. This measurement extends the data to four strange particle species
reported in \cite{xl1} with two strange resonances
the $\Sigma(1385)$ and the $K^*(892)^0$, and two neutral strange particles, 
the $\Lambda$ and the $K^0$. Strange resonances such as $K^*(892)^0$
and $\Sigma(1385)$ are particularly interesting to probe the earlier
stage of the collision and the subsequent evolution of the medium.
Indeed, these resonances are produced at sub-threshold energy and their productions
stemmed mainly from the fusion of pions and $K^+$ ($\Lambda$)~\cite{Bleicher:2002rx}, the latest
being produced at the beginning of the collision \cite{Hartnack:2001zs,hart21}. 
In addition, $K^*(892)^0$ and $\Sigma(1385)$ have short life time (4 and 5 fm/$c$, respectively)~\cite{rev}.
The measurement of  $K^*(892)^0$ and $K^0$ is detailed
in the present letter and their yield ratio is  
compared to the predictions of the UrQMD transport model.


The experiment was performed at the Heavy-Ion 
Synchrotron SIS of the GSI in Darmstadt by using an Al beam of kinetic energy of 
1.9$A$~GeV on an Al target. The target thickness was 567~mg/cm$^2$
and the average beam intensity was $8\cdot 10^5$ ions/s.
The FOPI detector is an azimuthally symmetric apparatus made of several 
sub-detectors which provide charge and/or mass determination over nearly 
the full solid angle. 
The central part of the detector is placed in a super-conducting solenoid and 
consists of a Central Drift Chamber (CDC) surrounded by a plastic scintillators 
(Barrel) for time of flight measurements. The forward part is composed of two walls 
of plastic scintillators (PLAWA and ZDD) and another drift chamber (Helitron) placed inside the
super-conducting solenoid. More details on the FOPI apparatus and its different sub-detectors
can be found in~\cite{Gobbi:1992hw, Ritman:1995td, Andronic:2000cx}.
The target was placed at 40 cm upstream of its nominal position
in order to cover the whole backward hemisphere \cite{mmxl}. 
Events are selected according to their centrality which is determined with the 
charged particle multiplicity measured in the CDC and in the PLAWA. 
The results presented here correspond to central collisions 
($\sigma_{geo}=315$ mb) representing about 20$\%$ of the total geometrical cross section.

For the present analysis, the $K^*(892)^0$ resonance (called $K^{*0}$ in the following) 
was reconstructed from its decay products in the channel $K^+ \pi^-$ (branching ratio 66\% \cite{rev}). 
Pions are measured in the CDC ($23^\circ < \theta_{lab} < 114^\circ$) and are identified by their mass 
determined by the correlation between magnetic rigidity and specific energy loss. 
$K^+$ are identified by matching the CDC and the surrounding ToF-barrel. This restricts the measured 
phase space of $K^+$ to $32^\circ < \theta_{lab} < 57^\circ$. More details on $K^+$ identification with the
FOPI detector are provided in \cite{Crochet:2000fz,Wisniewski:2001dk}.
Long-lived neutral strange particle like the $K^0_S$ are identified via their charged decay particles
in the CDC. A complete description of the $K^0_S$ reconstruction method can be found in~\cite{mmxl}.

\begin{figure}[!th] 
\includegraphics[width=6.5cm]{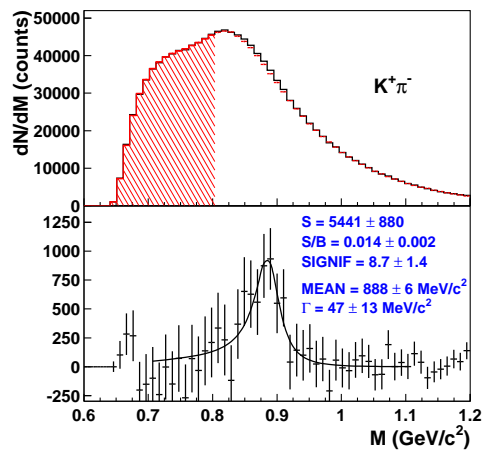} 
\caption{\label{fig:minv1}(Color online) Invariant mass spectra of $K^+\pi^-$ pairs.
The solid histogram and the dashed lines denote the data and the 
scaled mixed-event background, respectively (upper panel). The lower panel 
shows the signal after background subtraction. The following characteristics of 
the signal are shown: number of counts in the signal (S), 
signal-to-background ratio (S/B) and significance (SIGNIF = $S/\sqrt{S+B}$). 
The parameters extracted from the fit to the data 
(mean mass value (MEAN) and the width ($\Gamma$)) are also reported.} 
\end{figure}

The measurement of $K^{*0}$ corresponds to deep sub-threshold 
production (800 MeV) since the threshold  energy to create $K^{*0}$ in elementary reaction is 2.75 GeV. 
Because of its strong decay into $K^+$ and $\pi^-$, the short life time of this resonance does not allow us
to disentangle its decay vertex from the collision vertex. Therefore, the invariant 
mass analysis consists of correlating primary kaons and pions. 
The invariant mass plot of reconstructed $K^{*0}$ is shown in Fig.~\ref{fig:minv1}.
In order to enhance the signal-to-background ratio, a set of conditions was imposed:
\begin{itemize}

\item[(i)]   a cut on $K^+$ momentum ($p^{lab} \le 0.68$~GeV/c) is applied to reduce the
             contamination from fast pions and protons; 

\item[(ii)]   a cut on the $\pi^-$ transverse momentum ($p_{t}^{ lab}>0.185$~GeV/c) is used
             to reject pions which are spiraling in the CDC;

\item[(iii)] a cut on the distance of closest approach (DCA) between a track and the
             collision vertex in the transverse plane (DCA$_{K^+/\pi^-} <1$~cm) was applied
             to suppress decay products outside the target;

\item[(iv)]  a condition was imposed on the minimum number of activated 
             wires forming a track (hits) in the CDC (30 for $K^+$ and 40 for $\pi^-$)
	     for a higher quality of track reconstructions;

\item[(v)]   the following cut was applied for the systematic study of the signal (see text for more details): 
             $ 160 ^\circ < \theta^*_{K^+} + \theta^*_{\pi^-} < 200 ^\circ$,
	     $\theta^*$ being the angle between the momentum vector of a decay product in the 
             reference frame of $K^{*0}$ and the momentum vector of $K^{*0}$ in the reference 
             frame of the collision. Its relevance is described in \cite{Hyppolite_star}.

\end{itemize}
 
The combinatorial background is obtained with the event-mixing method. 
The two decay particles ($K^+$ and $\pi^-$) are taken from two 
different events characterized by the same particle multiplicity in the CDC. 
In addition, the two events are aligned to the reaction plane in order 
to have the same reference system for both particles. The reaction plane is estimated 
event by event utilizing the standard transverse momentum procedure 
proposed in~\cite{Danielewicz:1985hn}.
The combinatorial background was normalized 
on the left and right side of the nominal mass of the $K^{*0}$. For either sides 
no difference was found on the signal characteristics (mean, width and counts). Figure 1
shows the normalization on the left side (hatched region in upper panel of Fig.~\ref{fig:minv1}). 
The shape of the resulting mixed-event background describes the combinatorial 
background and is indicated by dashed lines in Fig.~\ref{fig:minv1} (upper panel)
where the statistical error bars are smaller than the line thicknesses.
After background subtraction (lower panel of Fig.~\ref{fig:minv1}), within the 
extracted width (47 MeV), about 5400 $K^{*0}$ are reconstructed 
from the analysis of 290 million events. The fitted function used to extract 
the number of counts of the signal is a Relativistic Breit-Wigner function 
modified by a Boltzmann factor to account for modification of the populated 
phase space in an hadronic heat bath \cite{Adams:2005}.
The mean mass and width extracted from the fit are in a good agreement, 
within statistical errors, with the values reported by the Particle Data Group~\cite{rev}.
The losses due to decay, acceptance and reconstruction efficiency have to be 
corrected in order to extract the particle yield. 
These corrections are determined by means of GEANT simulations 
modeling the full detector response. 
The phase space distribution of $K^{*0}$ was generated using the 
Siemens-Rasmussen formula~\cite{sie} which describes a radially expanding equilibrated source 
characterized by a temperature $T$ and a radial expansion velocity $\beta$.	
The choice of the values of these parameters 
($T=90~{\rm MeV}$, $\beta=0$ and $0.3$) is imposed by previous analysis
measurements~\cite{Hong:1997mr,xl1} and the values of the radial flow are
detailed below. The $K^{*0}$ particles are embedded into an heavy-ion event
calculated with the IQMD model~\cite{Hartnack:1997ez}.
The resulting output
is subject to the same reconstruction procedure as for the experimental data. 
The distributions of simulated and measured event for all relevant quantities were 
carefully compared and found to be in good agreement \cite{xl,mm1}. Finally, the reconstruction 
efficiency is determined by computing the ratio of reconstructed particles 
in the simulation to those initially embedded into the background events.
The systematic error on the $K^{*0}$ yield was evaluated in two steps.
First the reconstruction efficiency was estimated from simulated
$K^{*0}$ with two values of the radial flow velocity
($\beta=0$ and $\beta=0.3$). Then the $K^{*0}$ signals were reconstructed under 
different sets of conditions on the relevant quantities used to perform the reconstruction,
as for example, different cuts on $K^+$ and $\pi^-$ momentum. 
The use of a new variable in FOPI analysis like the $\theta^*$ \cite{Hyppolite_star}, 
variations of histogram characteristics (bin-size) and different normalization areas used to 
fit the background were also taken into account for the systematic calculation. 

\begin{figure}[!th]
\includegraphics[width=6.5cm]{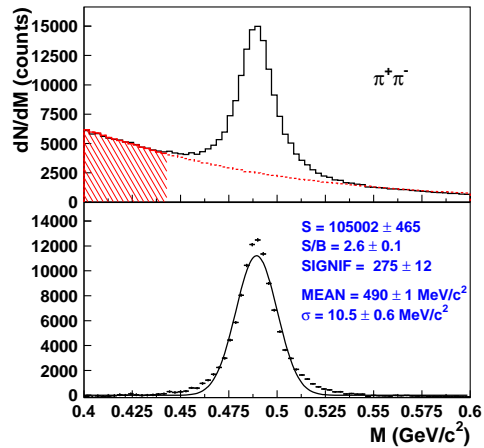}
\caption{\label{fig:minvk0}(Color online) Invariant mass distribution of $\pi^+\pi^-$ pairs.
The solid histogram and the dashed lines denote the data and the 
scaled mixed-event background, respectively (upper panel).
The lower panel shows the signal after background subtraction.
See caption of  Fig.~\ref{fig:minv1} for the abbreviations.}
\end{figure}
\begin{figure}[!th]
\includegraphics[width=4.25cm]{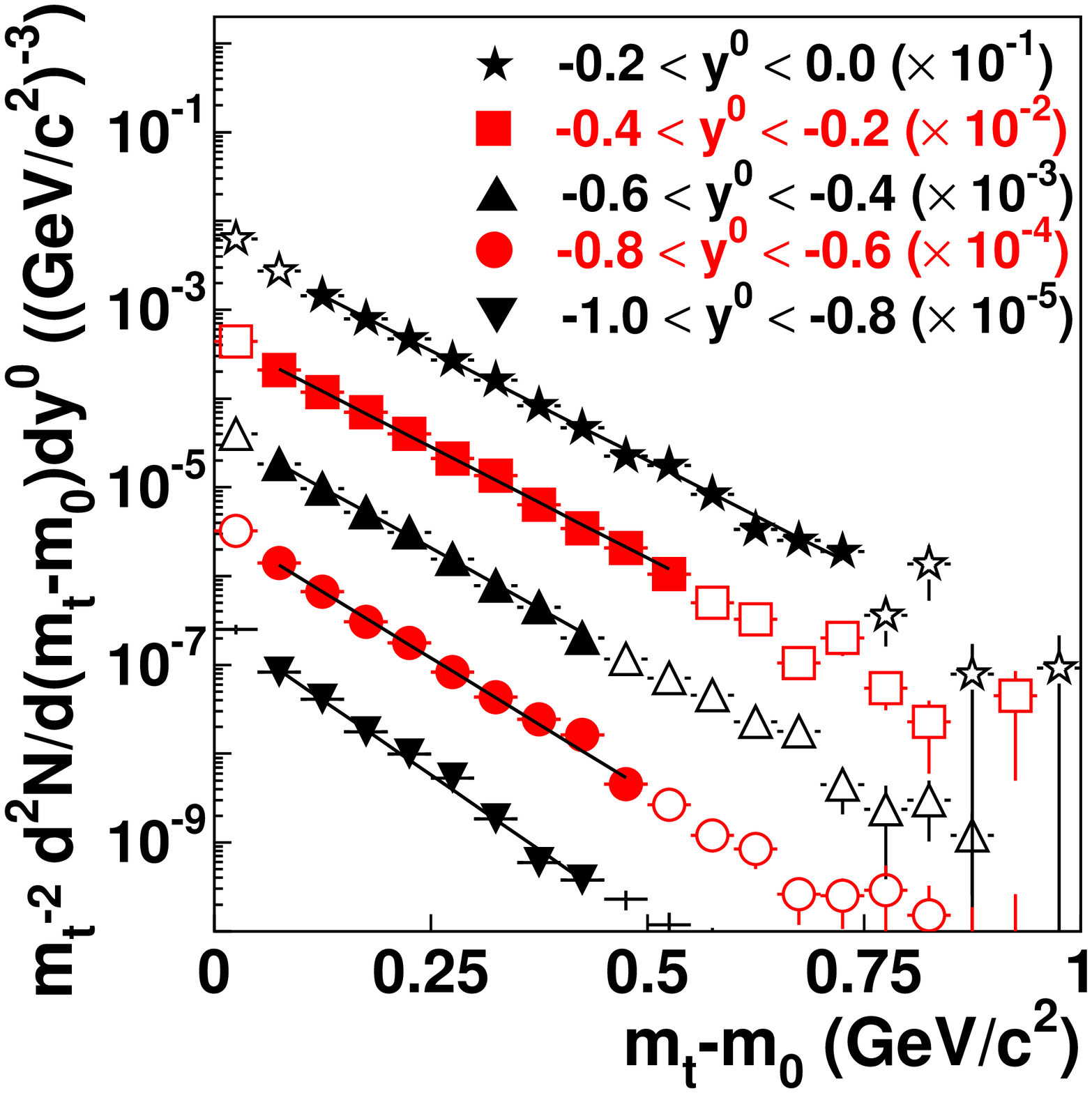}
\includegraphics[width=4.25cm]{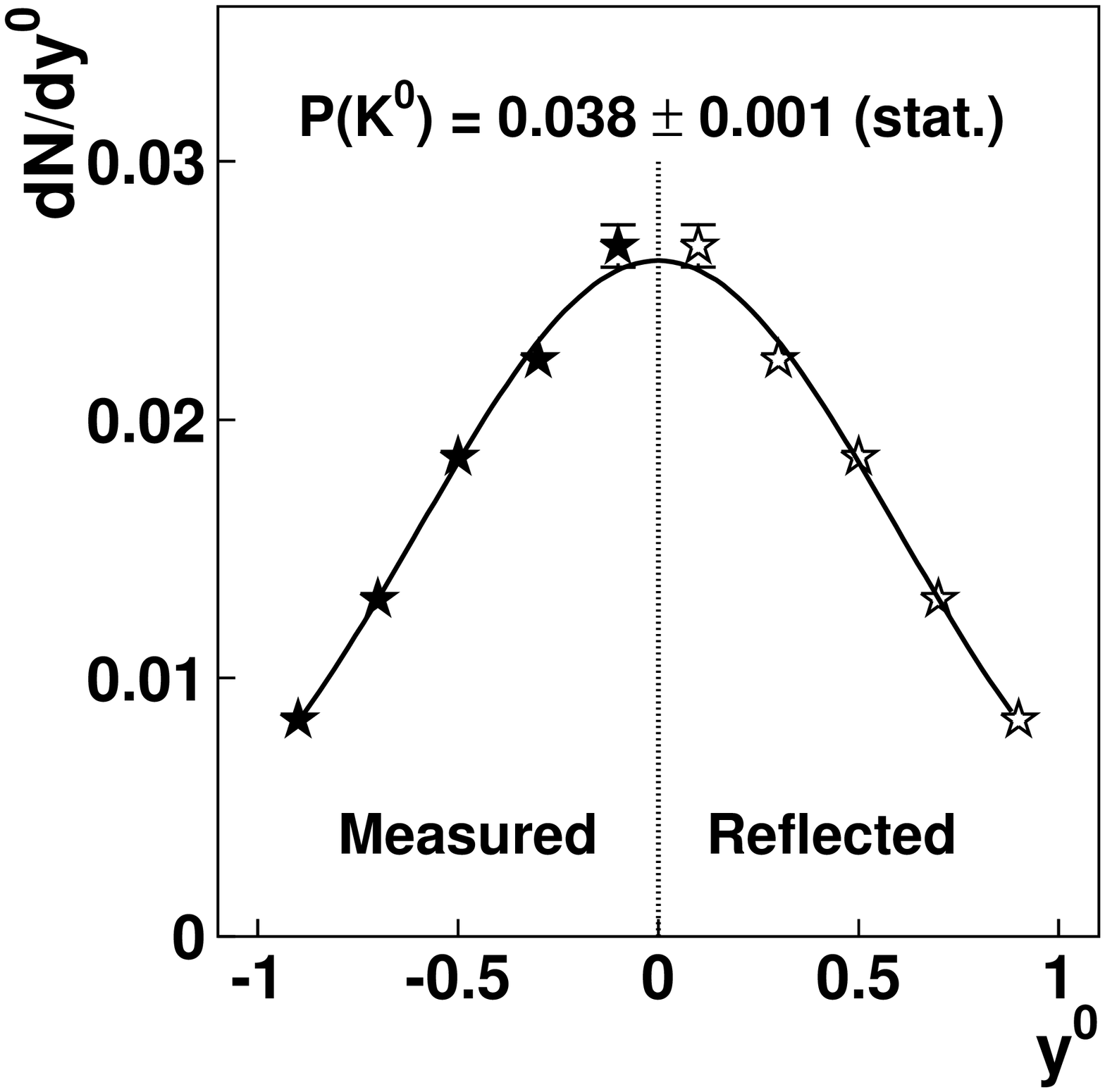} 
\caption{\label{fig:islp} (Color online)
Left: Transverse mass distributions of $K^0_S$ mesons for different rapidity windows
after efficiency corrections.
The lines represent the Boltzmann function fits. The range of the data 
used to perform the adjustment are shown by full symbols. The values in parentheses 
denote scaling factors used for the plot.
Right: Rapidity density distribution for $K^0$ where the open symbols are data points 
reflected with respect to mid-rapidity. The curve represents a Gaussian fit
from which the total yield is extracted (see text for more details).
}
\end{figure}

$K^0_S$ were reconstructed in the $\pi^+ \pi^-$ decay channel (branching
ratio 69.2\% \cite{rev}) from its four momenta at the intersection point of the two 
pions tracks (Fig.~\ref{fig:minvk0}). 
The combinatorial background was evaluated, as for the $K^{*0}$ 
case, with the event mixing method. After background subtraction, a signal of about 10$^5$ counts
within a $\pm 2\sigma$ window is extracted.
 The large statistics of reconstructed candidates allows to perform a bi-dimensional ($p_t$, $y^0$)
efficiency correction in order to extract the yield of $K^0$, where $y^0$ is the reduced rapidity
($y^0 = (y^{lab}-y^{cm})/y^{cm}$). The procedure used to 
simulate $K^0_S$ signals and Al+Al collisions is the same as the one used for the $K^{*0}$ particle. 
Decay products of $K^0_L$ cannot be measured with the FOPI apparatus. 
Since $K^0_S$ and $K^0_L$ are produced in equal amount (neglecting CP violation), 
the yield of $K^0$ corresponds to two times the one of $K^0_S$. 
The measured transverse mass spectra ($m_t=\sqrt{p_t^2+m_0^2}$) of  $K^0_S$ 
after efficiency correction are shown on the left side of Fig .~\ref{fig:islp}. 

In order to extract the yield,
we fit these distributions with a Boltzmann like function: 
\begin{equation*}
 \frac{1}{m^2_t}\frac{d^2N}{d(m_t-m_0)dy^{0}} = A \cdot exp \frac{-(m_t-m_0)}{T_B}  
\end{equation*}
within a narrow window of rapidity $dy^{0}$. 
One can notice that, in the logarithmic representation, the measured transverse mass distributions 
exhibit a linear decrease with increasing $m_t-m_0$ and are described reasonably well by the Boltzmann 
function. 
The latter allows to extract the rapidity density distributions $dN/dy^{0}$ 
by integrating the fitted function from $m_t-m_0$ = 0 to $\infty$.  
The symmetry of the colliding system allows us to reflect the rapidity spectrum around mid-rapidity.
The result, displayed in the right panel of Fig.~\ref{fig:islp}, is fitted 
with a Gaussian function after applying correction factors by taking into account the branching ratio
of $K^0_S$ ($\times$100/69) and the unmeasured $K^0_L$ ($\times$2). The integration of the Gaussian distribution
allows to extract the $K^0$ yield per central (315 mb) collisions: P($K^0$) = 0.038 $\pm$ 0.001(stat.) $\pm$ 0.004 (syst.).
~The~systematic~errors~of~the~measurement~are obtained by using the same 
procedure applied for the $K^{*0}$ particles.
Finally, the $K^{*0}$ to $K^0$ ratio is found to be:
\begin{equation*}
\label{yield}
\frac{\mathrm{P}(K^{*0})}{\mathrm{P}(K^{0})} = 0.0315\pm 0.006 (\mathrm{stat.})\pm 0.012 (\mathrm{syst.}).
\end{equation*}
Since for the $K^{*0}$ resonance the efficiency correction is applied with a global factor due to the low statistics
of the signal, such a ratio presents the advantage to cancel out in some extent the influence of 
the momentum distribution chosen for the simulated signals. 
The dependence on rapidity of the Boltzmann temperature ($T_B$) of $K^0_S$
is found to be in good agreement with an isotropically emitting source ($T_B=T_{eff}/cosh(y_{c.m.})$) \cite{mmxl}, where y$_{c.m.}$ 
is the rapidity in the nucleus-nucleus center of mass. 
For the mid-rapidity region, the inverse slope reaches $T_{eff}=89 \pm 1$(stat.) MeV and is consistent with the choice
of $T=90$ MeV used to generate simulated signals.

These measurements complete the study of strangeness production in Al+Al collisions at 1.9$A$~GeV
as presented in \cite{xl1}. A summary of our experimental results is shown in Tab.~\ref{tab-1}.
The measured yield ratios are compared to the predictions by
the UrQMD transport model \cite{Bleicher:2002rx,urqmd1,urqmd2,vogel}.
\begin{table}[!h]
\begin{tabular}{|c|c|c|}
 \hline
  Yield ratios & Data & UrQMD  \\
  \hline
  \hline
  $p/\pi^-$ & 5.3 $\pm$ 2.4 & 6.9 \\
  \hline
  $(\Lambda+\Sigma^0)/\pi^-$ & 0.017 $\pm$ 0.007 & 0.018 \\
  \hline
  $K^0/(\Lambda+\Sigma^0)$ & 0.66 $\pm$ 0.13 & 0.78 \\
  \hline
  $(\Sigma^{*+}+\Sigma^{*-})/(\Lambda+\Sigma^0)$ & 0.125 $\pm$ 0.042 & 0.177\\
  \hline
  $K^{*0}/K^{0}$ & 0.032 $\pm$ 0.013 & 0.024 \\
  \hline
\end{tabular}
\caption{\label{tab-1} Experimental ratios and predictions from
UrQMD model. The error of the experimental results
corresponds to the quadratic sum of statistical and systematic errors.}
\end{table}  
The UrQMD model allows to simulate heavy ion reactions according to a N-body theory 
on an event by event basis. It is the only QMD-based model 
which allows the production of $\Sigma^*$ and $K^{*}$ at SIS energy \cite{urqmd1,urqmd2}.
In this model, the Fermi motion (which is an important ingredient for particle
production below threshold) is treated in the following way. The initial momenta of 
the nucleons are randomly chosen between 0 and the local Thomas-Fermi momentum \cite{urqmd1}.
The parameters used for the latter are set to their default values as reported in \cite{user_guide}. 
The model predictions are found to be in good agreement
with the experimental ratios and their uncertainties. It is important 
to note that there is no explicit in-medium modifications of particle properties 
in this transport code.
The good agreement deduced from the comparison of model predictions and data
could be explained by a low nuclear density reached during the collision in the light Al+Al system.
A detailed discussion concerning the $\Sigma^*$ resonance production is provided in \cite{xl1}.
The dominant $K^{*0}$ production process involved in the UrQMD model
is the fusion of kaons and pions (70\%). The other processes to create $K^{*0}$ 
involve reactions of non-strange resonances ($\Delta$ and $N^*$) with baryons. The time dependence of these 
reactions exhibits a maximum around 7.5 fm/$c$.
This coincides with the production time of $K^+$ as predicted in \cite{hart21}.
This imply that the kaon-pion fusion occurs directly after kaon production.


In summary, we have presented new results on the P($K^{*0}$)/P($K^0$) 
strange mesons ratio in Al+Al collisions at a beam energy of 1.9$A$~GeV.
For the first time, the strange resonance $K^{*0}$ was measured 800~MeV below its 
production threshold in elementary reactions.
The transverse mass spectra of $K^0_S$  were found to be consistent 
with a Boltzmann distribution. 
The experimental results on strange particles measured with the FOPI
apparatus were compared to the predictions of the UrQMD transport model.
The measurements are in good agreement with the model predictions where
the dominant process to produce $K^{*0}$ at sub-threshold energy is the fusion of kaons
and pions.
The measurement of the $K^{*}$ strange resonance in an
heavier systems at sub-threshold energy could bring new informations
on in-medium modification of particle properties at higher baryon density
that can be reached at SIS energies and at the future FAIR facility.
   


We are grateful to M. Bleicher and S.~Vogel for providing us model calculations and for
intensive discussions.
This work was supported by the German BMBF under Contract No.~06HD154, 
by the the National Research Foundation of Korea (NRF) under grant No.
2009-0070676, by the mutual agreement between GSI and IN2P3/CEA,
by the Polish Ministry of Science and Higher Education under grant No. DFG/34/2007, 
by the Hungarian OTKA under grant No. 71989, within the Framework of the 
WTZ program (Project RUS 02/021), by DAAD (PPP D/03/44611) and by DFG 
(Projekt 446-KOR-113/76/04). We have also received support by the European 
Commission under the 6th Framework Program under the Integrated Infrastructure on:
 Strongly Interacting Matter (Hadron Physics), Contract No.~RII3-CT-2004-506078.


\end{document}